\documentclass{ws-procs961x669}            

\usepackage{xcolor}

\newcommand{\pwn}{PWNe}

\newcommand{\numin}{\nu_{\rm min}}
\newcommand{\numax}{\nu_{\rm max}}
\newcommand{\cn}[1]{Ref. \citenum{#1}}
\newcommand{\swift}{\textit{Swift}}

\begin{document}
\title{Applying models of pulsar wind nebulae to explain X-ray plateaux following short gamma-ray bursts}

\author{L. C. Strang$^*$ and A. Melatos}

\address{School of Physics, University of Melbourne,\\
  Parkville, VIC 3010, Australia\\
$^*$E-mail: lstrang@student.unimelb.edu.au}

\address{Australian Research Council Centre of Excellence in Gravitational-wave Discovery (OzGrav), \\School of Physics, University of Melbourne,\\
  Parkville, VIC 3010, Australia\\
$^*$E-mail: lstrang@student.unimelb.edu.au}

\begin{abstract}
 Many short Gamma-Ray Bursts (sGRBs) have a prolonged plateau in the X-ray afterglow lasting up to tens of thousands of seconds.
 A central engine injecting energy into the remnant may fuel the plateau.
 A simple analytic model describing the interaction of the magnetized relativistic wind from a rapidly-rotating magnetar with the surrounding environment can reproduce X-ray plateaux and instantaneous spectra.
 The model is analogous to classic, well-established models of young supernova remnants and applies the underlying physics to sGRB remnants.
 The light curve and spectra produced by the model are compared to observations of GRB 130603B.
 The spectra are also used to estimate parameters of the magnetar including its poloidal field strength and angular frequency.
 If combined with a gravitational wave signal, this model could provide insight into multimessenger astronomy and neutron star physics. 
\end{abstract}


\bodymatter
\section{Introduction}
Many short gamma-ray bursts (sGRBs) display long-lived emission in the X-ray band. 
A subset of the X-ray afterglows are ``canonical'' afterglows with three components: an initial decay, a flat plateau lasting from 10 s -- $10^5$ s, and a final decay (typically $\sim t^{-2}$) \cite{Zhang2006,Nousek2006}.
The luminosity and duration of the plateau are correlated, with brighter plateaux ending sooner\cite{DainottiCardone2008}.
A similar phenomenon is observed in long gamma-ray bursts (LGRBs), but the two populations are statistically distinct and likely have different progenitors for the afterglow as well as the prompt emission\cite{Dainotti2010,RowlinsonGompertz2014, rea2015constraining}.

Binary neutron star coalescence has been confirmed as a progenitor of sGRBs \cite{Abbott2017a,Abbott2017b}, suggesting the evolution of the post-merger remnant dictates the evolution of the afterglow.
X-ray plateaux have inspired several models including fireballs (with\cite{Rees1998,DainottiLenart2021} and without\cite{Piran1999} energy injection), fall-back accretion onto a black hole\cite{KumarNarayan2008}, and ongoing energy injection via a central engine\cite{DaiLu1998}, commonly assumed to be a millisecond magnetar\cite{ZhangMeszaros2001}.
In the latter scenario, the rotational energy of the magnetar is converted to X-rays via an unknown dissipative process, perhaps involving a relativistic wind\cite{DaiLu1998,Zhang2001,ZhangMeszaros2001,fan2005late,Gompertz2013,RowlinsonOBrien2013,Lasky2017}. 
Previous models have considered the evolution of a magnetar surrounded by a shroud of optically-thick ejecta material \cite{Metzger2014,Siegel2016,Yu2013} or the production of X-rays via radiative losses from interactions with the surrounding environment \citep{2018ApJ...869..155S,sarin2020interpreting}.

In this work, we present results from \cn{StrangMelatos2019} and \cn{StrangMelatos2021} exploring a millisecond magnetar central engine through the lens of classic models of pulsar wind nebulae (PWNe), also known as plerions.
The term plerion is used throughout \cn{StrangMelatos2019} and \cn{StrangMelatos2021}.
Here, we use the term \pwn \ throughout to emphasize the analogy to supernova remnant models such as \cn{PaciniSalvati1973} (although technically, we are discussing a magnetar wind nebula and not a pulsar wind nebula). 
In the PWNe model, the X-ray plateau is caused by a magnetized bubble of electrons fuelled by the relativistic wind of the magnetar. 
We summarize the model presented in \cn{StrangMelatos2019} in Sec.~\ref{sec:model} and explore associated synchrotron light curves in Sec.~\ref{sec:lcs}.
In Sec.~\ref{sec:spectra}, we quote results from \cn{StrangMelatos2021} inferring parameters of the central engine using instantaneous spectra.
We conclude in Sec.~\ref{sec:conclusion}.
This paper borrows substantially from the form and content of \cn{StrangMelatos2019} and \cn{StrangMelatos2021}; among other things, it includes Fig. \ref{fig:corner} from \cn{StrangMelatos2021} unchanged.


\section{Pulsar Wind Model}
\label{sec:model}
If a neutron star survives the sGRB, it evolves under the same physics that dictates the early evolution of \pwn.
Classic one-zone models for \pwn \ provide an analytic estimate of the synchrotron luminosity of the remnant without specifying its detailed geometry \cite{PaciniSalvati1973,ReesGunn1974}.
In \cn{StrangMelatos2019}, an analogous model is developed and applied to X-ray plateaux.
In this scenario, the sGRB heralds the birth of a rapidly-rotating neutron star with a dipole, magnetar-strength external magnetic field with angular frequency $\Omega(t)$, initial angular frequency $\Omega_0$, and polar surface magnetic field strength $B_0$. 
Simultaneously, an isotropic, relativistic blast wave (described by the Blandford-McKee self-similar solution\cite{BlandfordMcKee1976}) detonates and sweeps up the surrounding interstellar medium into a spherical shell expanding at $v_{\rm s}$.

The neutron star spins down under pure magnetic-dipole braking and radiates energy at a rate 
\begin{equation}
  \label{eqn:lsd}
L_{\rm sd}(t) = L_0 \left( 1+\frac{t}{\tau}\right)^{-2}
\end{equation}
where $L_0\propto B_0^2\Omega_0^4$ is the initial spin-down luminosity and $\tau \propto B_0^{-2} \Omega_0^{-2}$ is the characteristic spin-down timescale.
Through analogy with classic models of \pwn , we assume the energy is extracted as a magnetized, relativistic electron-positron wind with velocity $v_{\rm w} \gg v_{\rm s}$.
The ratio of the Poynting flux to kinetic-energy flux in the wind, $\sigma$, is expected to be $\gtrsim 1$ (i.e. the wind is Poynting-flux dominated) near the star and $\ll 1$ (i.e. kinetic-energy flux dominated) at distances far beyond the co-rotation radius $c/\Omega(t)$ ($\sigma \approx 10^{-3}$ for the Crab\cite{KennelCoroniti1984crab,MelatosMelrose1996,BogovalovKhangulyan2008}).

As in \pwn, a reverse shock forms at radius $r_{\rm rs}$ where the ram pressure of the wind $P_{\rm ram}(r)$ balances the external static pressure $P_{\rm stat}(r)$\cite{ReesGunn1974}.
As the shock propagates into the wind, the electrons are shocked into a power-law energy distribution $\propto E^{-a}$.
Reference \citenum{StrangMelatos2019} models the electron population with a spherical, homogeneous bubble characterized by its properties at $r_{\rm rs}$ (i.e. at $r = \dot{r}_{\rm rs} t$ for constant $\dot{r}_{\rm rs}$).
Under this assumption, the fresh electrons in the energy range $\left[E_{-0}, E_{+0}\right]$ are injected into the bubble at a rate
\begin{equation}
  \label{eqn:inj}
  \dot{N}_{\rm inj} = \frac{L_{\rm sd}(t) E^{-a}(a-2)}{\left(1+\sigma\right)\left(E_{-0}^{2-a}-E_{+0}^{2-a}\right)}.
\end{equation}
The bubble expands at a rate $\dot{r}_{\rm rs}$, so electrons cool according to adiabatic expansion at a rate
\begin{equation}
  \label{eqn:adiabatic}
  -\left.\frac{{\rm d}E}{{\rm d}t}\right|_{\rm adiabatic} = \frac{E}{t}.
\end{equation}
Given a magnetic field in the bubble $B(t)$, the electrons also cool via synchrotron radiation at a rate
\begin{equation}
  \label{eqn:synchcool}
  -\left.\frac{{\rm d}E}{{\rm d}t}\right|_{\rm synch} = c_sE^2B(t)^2
\end{equation}
where $c_s = \mu_0e^4/9\pi c^3m_e^4$.
Two magnetic field configurations are presented in \cn{StrangMelatos2019}.
In the first case, henceforth model A, $B(t)$ is taken to be the far-field extension of the stellar dipole field at radial distance $\dot{r}_{\rm rs}t$ (see Eq.~(9) in \cn{StrangMelatos2019}).
In the second case, henceforth model B, $B(t)$ is taken to be an arbitrary, constant $B$.
Model B serves two purposes.
One, it allows the \pwn \ model (including magnetar spin down, electron injection, and cooling mechanisms) to be assessed in general terms, even if the specific magnetic field structure in model A is a poor approximation to reality.
Two, model B decouples $B(t)$ and the central engine parameters $B_0$ and $\Omega_0$ (which also govern $\dot{N}_{\rm inj}$), allowing for the possibility that $B(t)$ may be affected by components of the post-merger environment other than the central engine.

The number density of electrons per unit energy $N(E,t)$ can be found via the time-dependent, inhomogeneous partial-differential equation 
\begin{equation}
  \label{eqn:PDE}
  \frac{\partial N(E,t)}{\partial t} = \frac{\partial }{\partial E}\left[ \left.\frac{{\rm d}E}{{\rm d}t}\right|_{\rm adiabatic} + \left.\frac{{\rm d}E}{{\rm d}t}\right|_{\rm synch}\right] + \dot{N}_{\rm inj}
\end{equation}
where adiabatic cooling is defined as in Eq.~(\ref{eqn:adiabatic}), synchrotron cooling is defined in Eq.~(\ref{eqn:synchcool}), and electron injection $\dot{N}_{\rm inj}$ is defined in Eq.~(\ref{eqn:inj}).
Two analytic Green's function solutions (corresponding to models A and B) are found in \cn{StrangMelatos2019} and integrated to obtain $N(E,t)$.

\section{Light Curves}
\label{sec:lcs}

In this section, we highlight results from \cn{StrangMelatos2019} comparing the X-ray synchrotron light curves predicted by the model in Sec. \ref{sec:model} to those observed by the Neil Gehrels \swift \ telescope\cite{GehrelsChincarini2004,EvansBeardmore2007,EvansBeardmore2009}.

The synchrotron spectrum of an electron with energy $E_{\rm c}$ may be approximated as a Dirac-delta function centred at the characteristic synchrotron frequency $\nu_{\rm c} \propto E_{\rm c}^2 B(t)$.
The source frame synchrotron luminosity in the range $\left[\numin, \numax \right]$ is 
\begin{equation}
\label{eqn:lsyn}
L_{\numin - \numax} = \int_{\numin}^{\numax} {\rm d}\nu  c_s B(t)^2 E_{\rm c}^2 N(E_{\rm c},t).
\end{equation}
We define $L_{\rm X}$ as the source-frame luminosity in the band corresponding to the 0.3 keV -- 10 keV band in the lab frame, i.e. the luminosity observable by \swift.

\begin{figure}[ht]
  \centering
  \includegraphics[width=0.75\linewidth]{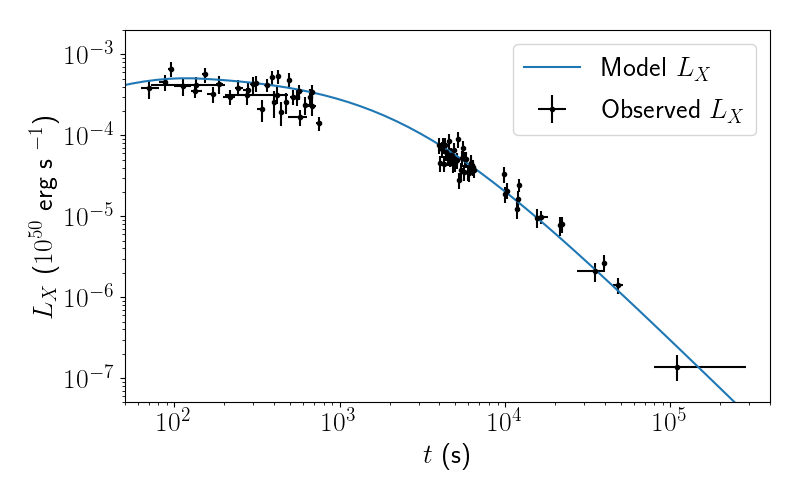}
  \caption{\label{fig:lc} X-ray luminosity in the source frame $L_{\rm X} \, (10^{50}{\rm erg \, s}^{-1})$ versus time $t \, {\rm (s)}$. Here ``X-ray'' refers to the 0.3 keV -- 10 keV band observable by \swift \, corrected to the source frame. Black crosses are observations of GRB 130603B by \swift \, and corrected to the source frame. The blue curve is the X-ray light curve predicted by the \pwn \ model with $B = 10 \, {\rm G}$, $B_0 = 4 \times 10^{15} \, {\rm G}$, $\Omega_0/2\pi = 140 \, {\rm Hz}$, and $a = 2.8$. Parameters chosen by hand, guided by results from \cn{StrangMelatos2021}.} 
\end{figure}

We use GRB 130603B as a representative example of X-ray plateaux following sGRBs because it has a known redshift\cite{Melandri2013} and has been inferred to spin down under magnetic-dipole braking\cite{Lasky2017}.
We display the observed light curve in Fig.~\ref{fig:lc} and overplot with a projected light curve from the \pwn \ model.
Black crosses are observations of GRB 130603B by \swift \, and corrected to the source frame.
The blue curve is the X-ray light curve predicted by the \pwn \ model with $B = 10 \, {\rm G}$, $B_0 = 4 \times 10^{15} \, {\rm G}$, $\Omega_0/2\pi = 140 \, {\rm Hz}$, and $a = 2.8$.
The parameters are chosen by hand, guided by results from \cn{StrangMelatos2021}.
The observed and predicted light curve are in broad agreement.

In some cases, such as GRB 090515, the plateau ends abruptly. 
In the context of the central engine model, this has been interpreted as the neutron star collapsing to a black hole and terminating energy injection\cite{RowlinsonOBrien2013} (we note, however, recent population-based arguments against this hypothesis\cite{BeniaminiLu2021}).
In the \pwn \ model, the collapse of the neutron star halts injection only but does not remove the existing population of electrons, so the afterglow persists for some time, given approximately by the synchrotron lifetime $t_{\rm synch} \sim 2\times 10^{-9}\left[B/(1 \, {\rm G})\right]^{-3/2}\left[\nu_c/(1\,{\rm keV})\right]^{-1/2} \, {\rm s}$.
In the X-ray band, and for the parameters in Figure 1, $t_{\rm synch}$ amounts to nanoseconds.

\subsection{Luminosity-time correlation}
The Dainotti correlation connects the plateau luminosity $L_{\rm p}$ and plateau duration $t_{\rm p}$ in GRB plateaux which states that brighter plateaux have shorter durations than dimmer plateaux\cite{DainottiCardone2008,DainottiWillingale2010,SultanaKazanas2012}.
This correlation has been studied extensively for LGRBs under a variety of assumptions\cite{DainottiPostnikov2016,DainottiBoria2018} (see \cn{Dainotti2019} for a thorough review on the correlation and its uses).

\begin{figure}[ht]
  \centering
  \includegraphics[width=0.75\linewidth]{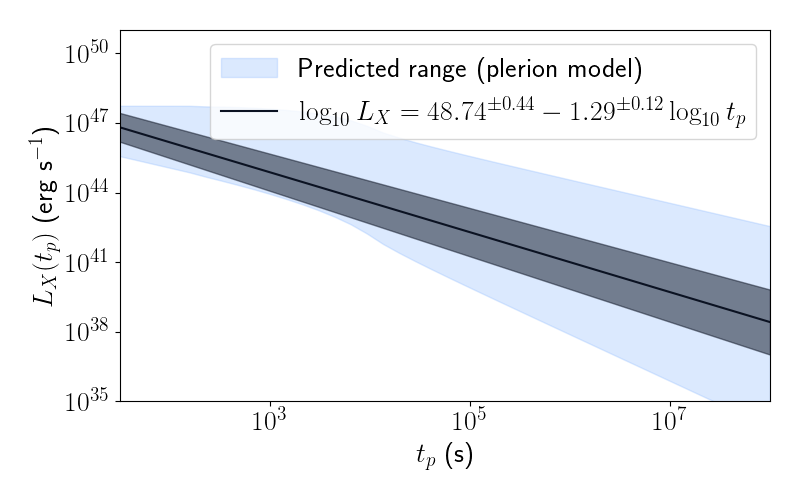}
  \caption{\label{fig:corrs} X-ray plateau luminosity in the source frame $L_{\rm p} = L_{\rm X}(t_{\rm p}) \, ({\rm erg \, s}^{-1})$  versus X-ray plateau duration $t_{\rm p} \, {\rm (s)}$. The black line and grey-shaded region correspond to the best-fit correlation from \cn{RowlinsonOBrien2013} from a sample of 159 GRBs. The light blue shaded region corresponds to the range of $\left(L_{\rm p}, \, t_{\rm p}\right)$ pairs generated by the \pwn \ model for $B = 5.0\times 10^{-1}\,{\rm G}$, $10^{12} \leq B_0/(1 \, {\rm G}) \leq 10^{16}$, and $\Omega_0/2\pi \leq 10^3 \, {\rm Hz}$. Model parameters were chosen based on results presented in \cn{StrangMelatos2021}. }
\end{figure}

\begin{figure}[ht]
  \centering
  \includegraphics[width=\linewidth]{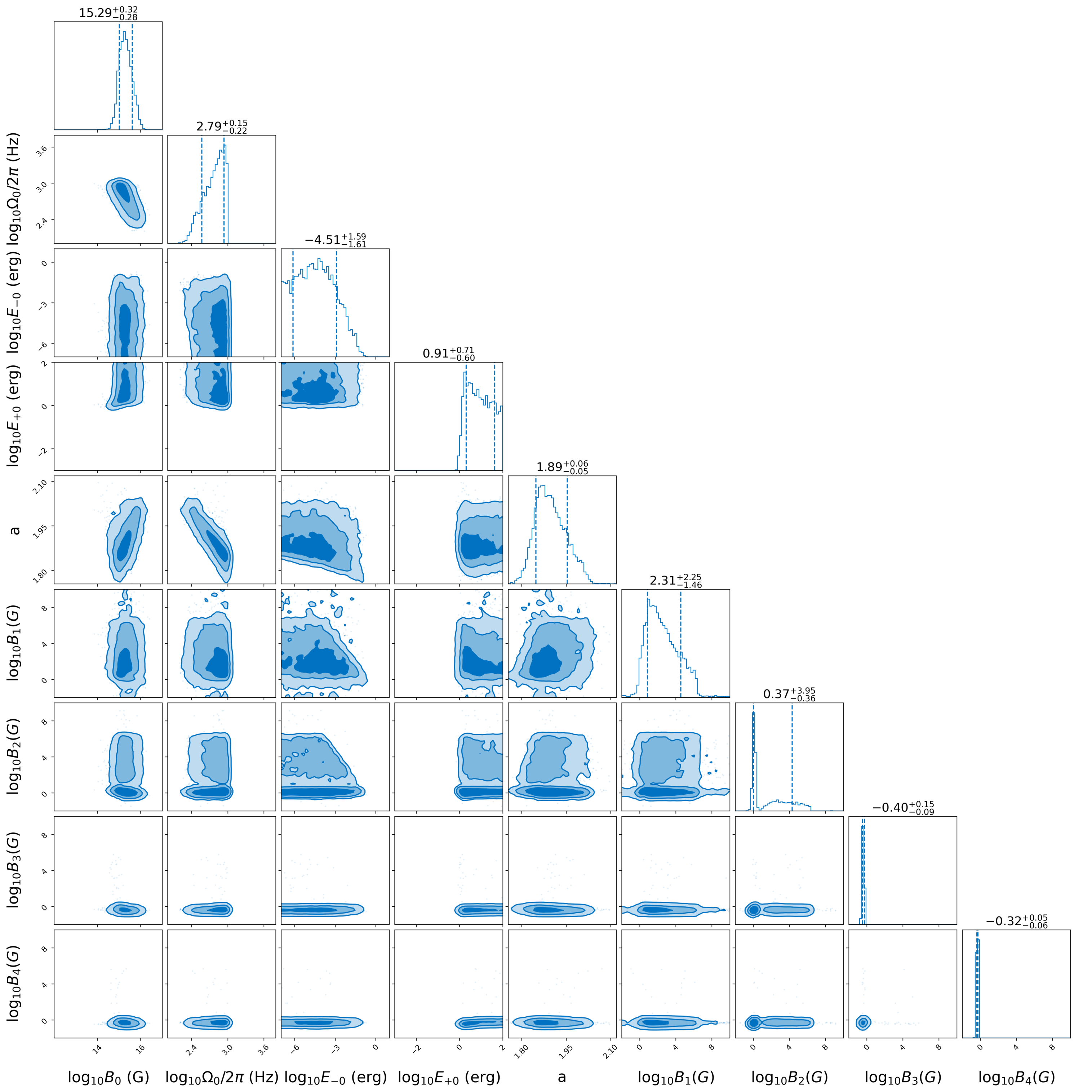}
  \caption{\label{fig:corner} Corner plot showing the posterior distribution obtained for four instantaneous spectra for GRB130603B for the parameters $\log_{10} B_0$ (G), $\log_{10}\Omega_0/2\pi$ (Hz), $\log_{10}E_{\pm 0}$ (erg), $a$, and $\log_{10}B_i \,(1\leq i\leq 4)$ (G). Figure reproduced from \cn{StrangMelatos2021}}
\end{figure}

The \pwn \ model presented in \cn{StrangMelatos2019} naturally reproduces the observed correlation for reasonable definitions of $L_{\rm p}$ and $t_{\rm p}$.
One reasonable choice is to define the X-ray plateau duration as $t_{\rm p} = \tau$ and the corresponding luminosity  as $L_{\rm p} = L_{\rm X}(t=\tau)$.
The light blue region in Fig.~\ref{fig:corrs} indicates the full range of $\left(L_{\rm p}, \, t_{\rm p}\right)$ pairs generated by the \pwn \ model for $B = 5.0\times 10^{-1}\,{\rm G}$, $E_{-0} = 2.5\times 10^{-2} \, {\rm erg}$, $10^{12} \leq B_0/(1 \, {\rm G}) \leq 10^{16}$, and $\Omega_0/2\pi \leq 10^3 \, {\rm Hz}$.
The black line and grey-shaded region corresponds to the best-fit correlation from \cn{RowlinsonOBrien2013} based on a sample of 159 sGRBs and LGRBs.
The observed correlation falls entirely within the range permitted by the \pwn \ model.

\section{Inferring parameters of the central engine}
\label{sec:spectra}
In this section, we review results presented in \cn{StrangMelatos2021} inferring $B_0$, $\Omega_0$, $E_{\pm 0}$, $a$, and (for model B) $B$ using the X-ray afterglow for six sGRBs of known redshift with an X-ray plateau.
We reproduce here the necessary physics to interpret the results and leave the details of the analysis to \cn{StrangMelatos2021}.

The source-frame synchrotron spectrum at time $t$ is 
\begin{equation}
  \label{eqn:specflux}
F_\nu(t, \nu) = N(E, t) \left.\frac{{\rm d}E}{{\rm d}t}\right|_{\rm synch}\frac{{\rm d}E}{{\rm d} \nu}.
\end{equation}
The synchrotron cooling rate and the characteristic synchrotron frequency depend on the magnetic field in the bubble, which is determined by $B_0$ and $\Omega_0$ in model A and by $B$ in model B.
The dependence on $N(E,t)$ indicates $F_\nu$ is dependent on $B$, $B_0$, $\Omega_0$, $E_{\pm 0}$, and $a$, so a single point-in-time spectrum is sufficient to infer posteriors on each parameter. 
For a full discussion of the results and method, see Sec. 4 in \cn{StrangMelatos2021}.
For both models, one has a neutron star with an approximately millisecond spin period ($\Omega_0/2\pi \lesssim 10^3 \, {\rm Hz}$) and a magnetar-strength poloidal magnetic field ($B_0 \sim 10^{15}\,{\rm G}$) i.e. a millisecond magnetar.
The correlations between $B_0$ and $\Omega_0$ reflect their relation to the spin-down luminosity $L_0 \propto B_0^2\Omega_0^4$ and $\tau \propto B_0^{-2}\Omega_0^{-2}$.
For model B, the posteriors on $B$ cover the range $10^{-1} \lesssim B/(1\,{\rm G}) \lesssim 1$.
This is much stronger than the magnetic field in the interstellar medium ($\sim 10^{-6} \, {\rm G}$) but smaller than the expected field advected outwards from the central object by the relativistic outflow (i.e. the magnetic field in model A).
For both models, $E_{-0} \sim 10^{-3} \, {\rm erg}$, which is consistent with a population of electrons accelerated in the magnetar's magnetosphere and injected into the magnetar wind with a radiation-reaction-limited Lorentz factor.

\subsection{Spectral evolution}
In this section, we review briefly results presented in \cn{StrangMelatos2021} using the spectrum of GRB 130603B at four separate epochs to infer $B_0$, $\Omega_0$, $E_{\pm 0}$, $a$, and $B$ for each epoch.
We label each epoch $t_i$ with corresponding magnetic field $B_i$ for  $i\in[1,4]$ and use log uniform priors such that $10^{-6} < B_i/(\rm 1\, G) < 10^6$.
For each epoch with flux data $F_{\nu_{\rm i}}$, flux uncertainty $\sigma_i$, and model flux $F_\nu(t_i,\nu)$ given by Eq. (\ref{eqn:specflux}), we define a Gaussian likelihood $P_i \propto {\rm exp}\left\{\left[F_{\nu_i} - F_\nu\left(t_i, \nu\right)\right]^2/\left(2\sigma_i^2\right)\right\}$ such that the priors on $B_0$, $\Omega_0$, $E_{\pm 0}$ and $a$ are the same at each epoch.
The analysis is performed on the joint likelihood $\Pi_i P_i$.

The median values of the posterior describe a millisecond magnetar with $B_0 \approx 2 \times 10^{15} \, {\rm G}$ and $\Omega_0/2\pi \approx 600 \, {\rm Hz}$, supplying the remnant with relativistic electrons with $a \approx 1.9$, $E_{-0} \approx 3.2 \times 10^{-5} \, {\rm erg}$, and $E_{+0} \approx 1 \, {\rm erg}$.
The median magnetic fields $B_i$ suggest the field drops at an average rate of $0.04 \, {\rm G \, s}^{-1}$ from $B_1 = 2 \times 10^2 \, {\rm G}$ at $t_1 = 643 \, {\rm s}$ to $B_4 = 5 \times 10^{-1} \, {\rm G}$ at $t_4 = 5735 \, {\rm s}$.
This is both slower and weaker than what is expected for the field in the termination shock of the wind in model A, which scales roughly as $B(t)  \propto B_0 t^{-2}$ for $t \gtrsim \tau$, if the wind expands at a constant, relativistic speed (as in \cn{StrangMelatos2019}).
The scale and temporal evolution of the $B_i$ indicate the time-dependent magnetic field in the bubble requires revision. 
Several mechanisms for this are discussed in \cn{StrangMelatos2021}, including varying the expansion rate of the bubble or choosing an entirely separate magnetic field configuration.

\section{Conclusion}
\label{sec:conclusion}
A millisecond magnetar engine based on classic models of \pwn \ can explain some of the observed features of canonical sGRB afterglows.
Plateau light curves are broadly consistent with the synchrotron output of an electron population $N(E,t)$ evolving under ongoing, power-law injection and adiabatic and synchrotron cooling. 
Such a model may be referred to as a plerion model (as in \cn{StrangMelatos2019} and \cn{StrangMelatos2021}) or, as a synonym, a \pwn \ model (as we do in this work).
The model is able to reproduce both the shape of the afterglow and correlations $L_p$--$t_p$.
Furthermore, the instantaneous spectra can be used to infer $B$, $B_0$, $\Omega_0$, $E_{\pm 0}$, and $a$ within the context of the model.
However, the model summarized here (originally in \cn{StrangMelatos2019}) is highly idealized and does not include the important effects of merger ejecta on the evolution of the remnant (considered in e.g. Refs. \citenum{Metzger2014,Siegel2016,Yu2013}).
Also, as presented in \cn{StrangMelatos2021}, the temporal evolution of the spectrum implies a magnetic field that is inconsistent with both a dipole field (model A) and a constant field (model B).
Future modeling improving on the work in \cn{StrangMelatos2019} and integrating it into existing models for sGRB remnants should ultimately be compared to a broad sample of sGRBs.

\section*{Acknowledgments}
This work makes use of data supplied by the UK Swift Science Data Centre at the University of Leicester.
Parts of this research were conducted by the Australian Research Council Centre of Excellence for Gravitational Wave Discovery (OzGrav), through Project Number CE170100004.
The work is also supported by an Australian Research Council Discovery Project grant (DP170103625).
\bibliographystyle{ws-procs961x669}
\bibliography{ws-pro-sample}{}

\end{document}